\documentclass[runningheads,a4paper]{llncs}

\usepackage{amssymb,fixltx2e}
\usepackage{amsmath} 
\usepackage{rotating}
\usepackage{float}
\usepackage{pgfplots} 
\usepackage{xcolor}
\usepackage{subcaption}
\usepackage{float}
\usepackage{pgfplots} 
\usepackage{tikz}
\usepackage{tikzscale}
\usepackage{graphicx}
\usepackage{algorithm}

\usepackage{bbm}

\usepackage[noend]{algpseudocode}
\usepackage{url}
\urldef{\mailsa}\path|{author1, author2, author3}@sinstitute.edu|
\newcommand{\keywords}[1]{\par\addvspace\baselineskip
\noindent\keywordname\enspace\ignorespaces#1}
\definecolor{bblue}{HTML}{4F81BD}
\definecolor{rred}{HTML}{C0504D}
\definecolor{ggreen}{HTML}{9BBB59}
\definecolor{ppurple}{HTML}{9F4C7C}

\renewcommand\textfloatsep{4pt}

\renewcommand\intextsep{4pt}

\begin{document}

\mainmatter  

\title{Multi-Task Learning for Extraction of Adverse Drug Reaction Mentions from Tweets}
\titlerunning{Extraction of Adverse Drug Reaction Mentions from Tweets}


%
%
\author{Shashank Gupta\inst{1} \and Manish Gupta\inst{1}\thanks{Author is also a Principal Applied Scientist at Microsoft} \and 
Vasudeva Varma\inst{1} \and Sachin Pawar\inst{2} \and Nitin Ramrakhiyani\inst{2} \and Girish Keshav Palshikar\inst{2}}
\authorrunning{Gupta et al.}
\institute{International Institute of Information Technology-Hyderabad, India\\
\email{shashank.gupta@research.iiit.ac.in}\\ \email{\{manish.gupta,vv\}@iiit.ac.in}
\and
TCS Research, Pune\\
\email{\{sachin7.p,nitin.ramrakhiyani,gk.palshikar\}@tcs.com}}

%
%

\maketitle

\begin{abstract}
Adverse drug reactions (ADRs) are one of the leading causes of mortality in health care. Current ADR surveillance systems are often associated with a substantial time lag before such events are officially published. On the other hand, online social media such as Twitter contain information about ADR events in real-time, much before any official reporting. Current state-of-the-art in ADR mention extraction uses Recurrent Neural Networks (RNN), which typically need large labeled corpora. Towards this end, we propose a multi-task learning based method which can utilize a similar auxiliary task (adverse drug event detection) to enhance the performance of the main task, i.e., ADR extraction. Furthermore, in absence of the auxiliary task dataset, we propose a novel joint multi-task learning method to automatically generate weak supervision dataset for the auxiliary task when a large pool of unlabeled tweets is available. Experiments with $\sim$0.48M tweets show that the proposed approach outperforms the state-of-the-art methods for the ADR mention extraction task by $\sim$7.2\% in terms of F1 score. 

\end{abstract}

\section{Introduction}
Estimates show that Adverse Drug Reactions (ADRs) are the fourth leading cause of deaths in the United States ahead of cardiac diseases, diabetes, AIDS and other fatal diseases\footnote{https://ethics.harvard.edu/blog/new-prescription-drugs-major-health-risk-few-offsetting-advantages}. Another study\footnote{http://bit.ly/2vaWF6e} conducted in the US reveals that $\sim$6.7\% of the hospitalized patients have a serious ADR, with a fatality rate of $\sim$0.32\%. Hence, it necessitates the monitoring and detection of such adverse events to minimize the potential health risks by having the relevant pharmaceutical companies issue appropriate warnings. Practically, clinical trials cannot investigate all settings in which a drug can be used, making it impractical to profile a drug's side effects before its formal approval. Typically, post-marketing drug safety surveillance (also called as pharmacovigilance) is conducted to identify ADRs after a drug's release. Such surveys rely on formal reporting systems such as Federal Drug Administration's Adverse Event Reporting System (FAERS)\footnote{http://bit.ly/2xnu7pE}. However, often a large fraction ($\sim$94\%) of the actual ADR instances are under-reported in such systems~\cite{hazell2005under}. Social media presents a plausible alternative to such systems, given its wide userbase. 
A recent study~\cite{freifeld2014digital} shows that Twitter has three times more ADRs reported as compared to FAERS. 

Earlier work in this direction focused on feature based pipeline followed by a sequence classifier \cite{nikfarjam2015pharmacovigilance}. More recent works are based on Deep Neural Networks \cite{cocos2017deep}. Deep learning based methods~\cite{collobert2011natural,lecun2015deep} typically rely on the presence of a large annotated corpora, due to their large number of free parameters. Due to the high cost associated with tagging ADR mentions in a social media post and limited availability of labeled datasets, it is hard to train a deep neural network effectively for such a task. In this work, we attempt to address this problem and propose two novel multi-task learning setups which utilize similar tasks to effectively augment the rather limited existing datasets for ADR extraction.   

Multi-task learning works on the basic premise that auxiliary tasks can be utilized to improve performance of the main task by exploiting the correlations between them \cite{evgeniou2004regularized}. Adverse drug event (ADE) detection is a task very similar to our original task of ADR mention extraction. The ADE detection problem deals with \textit{detecting} an adverse drug event from a social media post. We hypothesize that due to semantic similarities between the two tasks, they can be modeled together in a joint learning setup. We propose a multi-task learning setup with ADR extraction as the main task and ADE detection as an auxiliary task which complements the learning of our main task. Furthermore, we propose a novel weakly-supervised learning based method which exploits semi-supervised learning to augment the main task (ADR extraction) dataset and also works in parallel to automatically generate auxiliary task (ADE detection) dataset. 

To summarize, the main contributions of our work are: (1) We investigate the effect of adding an available auxiliary task (ADE detection) to the main task (ADR extraction) in a multi-task learning setup. (2) We propose a novel weakly-supervised and a semi-supervised learning based method to automatically generate auxiliary task dataset (ADE detection) and model it in a novel joint multi-task learning framework. (3) We perform experiments on two datasets to show the effectiveness of the proposed methods.

The remainder of the paper is organized as follows. In Section \ref{rel-work} we discuss the related work in the area of ADR extraction and Multi-task learning. In Section \ref{approach} we describe our proposed methods in detail. In Section \ref{experiments} and Section \ref{analysis}, we discuss in detail our experimental results and its analysis. Finally, Section \ref{conclusion} concludes our work with a brief summary.

\section{Related Work}\label{rel-work}
In this section, we review some of the existing work in the areas of ADR extraction and Multi-task learning.

\noindent \textbf{ADR Extraction:} Traditional methods for ADR extraction used linguistic features such as POS tags, word embedding features and word context features along with sequence classifiers like a linear-chain CRF~\cite{nikfarjam2015pharmacovigilance}. To avoid  time consuming feature engineering, recent works use deep learning approaches~\cite{collobert2011natural,DBLP:conf/emnlp/Kim14,lecun2015deep,DBLP:conf/emnlp/2015}. Cocos et al.~\cite{cocos2017deep} proposed a Long Short Term Memory (LSTM) based model with word embedding features to extract ADRs from Twitter posts. Stanovsky et al.~\cite{DBLP:conf/eacl/StanovskyGM17} proposed a LSTM based model where lexical word embeddings are augmented with Knowledge-Graph based embeddings. In their model, if a word has a lexical match with a Knowledge-Graph entity (e.g., DBPedia), its corresponding lexical word embedding is replaced by embedding learned through Knowledge graph based methods \cite{DBLP:conf/aaai/WangZFC14}.

\noindent \textbf{Multi-Task learning (MTL):} Previous works in Multi-task learning have explored the use of auxiliary tasks to improve the generalization performance of a main task \cite{DBLP:journals/jmlr/AndoZ05,DBLP:conf/nips/ArgyriouEP06,DBLP:conf/icml/Caruana93,evgeniou2004regularized}. In the context of deep neural networks, MTL has been successfully applied in the area of Natural Language Processing \cite{DBLP:conf/icml/CollobertW08,DBLP:journals/corr/LuongLSVK15} and Information Retrieval \cite{DBLP:conf/naacl/LiuGHDDW15}. These models work on the premise that multiple related tasks share common features which allows the model to share the statistical strengths between them. Sharing statistical strengths among different tasks also acts as an implicit regularizer, allowing the model to generalize better. Due to sharing of the model between tasks, MTL also effectively acts as an implicit data augmentation method, since the same model is exposed to the training data of multiple tasks. In this work, we exploit the data augmenter role of MTL to compensate for the lack of rich training data for the ADR extraction task using a single neural network based model. 

\section{The Proposed Multi-Task Learning Framework}\label{approach}
In this section, we start by defining the ADR extraction and ADE detection problems. Next, we propose a multi-task learning framework for ADR extraction. Finally, we propose a joint multi-task learning framework for both the tasks.

\subsection{Problem Definition}
\textbf{ADR Extraction:} Given a social media post in the form of a word sequence $x_1. ..., x_n$, predict an output sequence $y_1,....,y_n$ which indicates the presence/absence of the ADR mention, where each $y_i$ is encoded using standard sequence labeling encoding scheme such as the IO encoding similar to that used in~\cite{cocos2017deep}.

\noindent \textbf{ADE Detection:} Given a social media post in the form of a word sequence $x_1,...,x_n$, predict a single variable $y$, which indicates whether there is an occurrence of an ADE in the input social media post or not. It can thus be modeled as a binary classification problem.

\begin{algorithm}[t!]
\scriptsize
\caption{\scriptsize Multi-Task Learning for ADR Extraction}\label{alg:tml-adr}
\hspace*{\algorithmicindent} \textbf{Input} \hspace{0.3mm}  $N$: (No. of training examples / batch size) for ADR task \\
\hspace*{\algorithmicindent} \hspace*{\algorithmicindent} \hspace*{\algorithmicindent}   $M$: (No. of training examples / batch size) for ADE task \\
\hspace*{\algorithmicindent} \hspace*{\algorithmicindent} \hspace*{\algorithmicindent}   $\alpha$: $\frac{M}{N}$ \\
\hspace*{\algorithmicindent} \textbf{Output} \hspace{0.3mm} Model parameters: $\theta_{\text{Shared}}$, $\theta_{\text{ADR}}$, $\theta_{\text{ADE}}$
\begin{algorithmic}[1]
\State Initialize model parameters : $\theta_{\text{Shared}}$, $\theta_{\text{ADR}}$, $\theta_{\text{ADE}}$ randomly
\For{$epoch\gets 1, maxEpochs $}
\For{$i\gets 1, N$}
\For{$j\gets 1, \alpha$}
\State $X_{ADE}$, $Y_{ADE}$ = sample $(N(i-1)+j)^{th}$ batch from ADE training data
\State $L_{ADE}$ = ADE Loss($X_{ADE}$, $Y_{ADE}$) from Eq.~\ref{ade-loss}
\State Compute gradients for ADE loss, and update $\theta_{\text{Shared}}$, $\theta_{\text{ADE}}$
\EndFor
\State $X_{ADR}$, $Y_{ADR}$ = sample $i^{th}$ batch from ADR training data
\State $L_{ADR}$ = ADR Loss($X_{ADR}$, $Y_{ADR}$) from Eq.~\ref{adr-loss}
\State Compute gradients for ADR loss, and update $\theta_{\text{Shared}}$, $\theta_{\text{ADR}}$
\EndFor
\EndFor

\end{algorithmic}
\end{algorithm}

\subsection{Multi-Task Learning for ADR Extraction}
\label{sec:mtl}
Given the two tasks, ADR extraction and ADE detection, we first describe the modeling of each task individually and then discuss how to model them in a single setup.

\noindent \textbf{ADR Extraction:} We choose the model described in~\cite{cocos2017deep}, which is a fully supervised bi-directional LSTM (bi-LSTM) transducer trained on a manually annotated tweet corpus with word-level ADR mention annotation. Formally, given an input word sequence $x_1,....,x_n$, where $n$ is the maximum sequence length, a bi-LSTM transducer \cite{DBLP:journals/corr/abs-1211-3711} is employed to capture complex sequential dependencies. At each time-step $t$, the bi-LSTM transducer attempts to model the task as follows. 
\begin{eqnarray}
h_t = \text{bi-LSTM}(e_t, h_{t-1}) 
\end{eqnarray}
where $h_t \in \mathcal{R}^{(2\times d_h)}$, is the hidden unit representation of the bi-LSTM with $d_h$ being the hidden unit size. Since it is a concatenation of hidden units of a forward sequence LSTM and backward sequence LSTM, its overall dimension is $2 d_h$. $e_t$ is the embedding vector corresponding to the input word $x_t$ extracted from a pre-trained word embedding lookup table.
\begin{eqnarray}
y_t = \text{softmax}(W_1 h_t + b) 
\end{eqnarray}
\noindent where $y_t \in \mathcal{R}^{d_l}$, is the output vector at each time-step which encodes the probability distribution over the number of possible output labels ($d_l$) at each time-step of the sequence. $W_1 \in \mathcal{R}^{{d_l*d_h}}$ and $b \in \mathcal{R}^{d_l}$ are weight vectors for the affine transformation. Finally, the cross entropy loss function for the task is defined as follows.
\begin{eqnarray}
L_{\text{ADR}} = - \sum_{t=1}^{n} \sum_{i=1}^{d_l}  \hat{y_{t_i}} \log y_{t_i}\label{adr-loss}
\end{eqnarray}
\noindent where $\hat{y_t}$ is the one-hot representation of the actual label at time-step $t$.

\noindent \textbf{ADE Detection:} Given an input word sequence $x_1,....,x_n$, where $n$ is the maximum sequence length, similar to the ADR Extraction model, a bi-directional LSTM transducer (bi-LSTM) is employed to model the sequential nature of the dataset. The LSTM transducer acts as a feature extractor in this case, which is followed by an average-pooling layer to generate a fixed-size vector representation of the input sentence followed by the classification loss function. Formally, the ADE detection model is defined as follows.
\begin{eqnarray}
h_t = \text{bi-LSTM}(e_t, h_{t-1}), \ \ \ h = \frac{1}{n} \sum_{t=1}^{n} h_t, \ \ \ \ y = \text{softmax}(W_2 h + b_1)
\end{eqnarray}
\noindent where $h_t$ is similar to the one defined for the ADR task. $h \in \mathcal{R}^{(2*d_h)}$ is the average-pooled fixed size representation of the input sequence. $y \in \mathcal{R}^{2}$ is the output vector which encodes the probability distribution over the binary choice, with $W_2 \in \mathcal{R}^{{2 * (2d_h)}}$ and $b_1 \in \mathcal{R}^{2}$, the corresponding weight vectors. Finally, the loss function for the task is the cross-entropy loss defined as follows.
\begin{eqnarray}
L_{\text{ADE}} = -  \sum_{i=1}^{2}  \hat{y_i} \log y_i\label{ade-loss}
\end{eqnarray}
\noindent where $\hat{y}$ is the one-hot representation of the actual label for the input sentence.

\begin{figure}[t!]
\centering
  \includegraphics[width=0.7\linewidth]{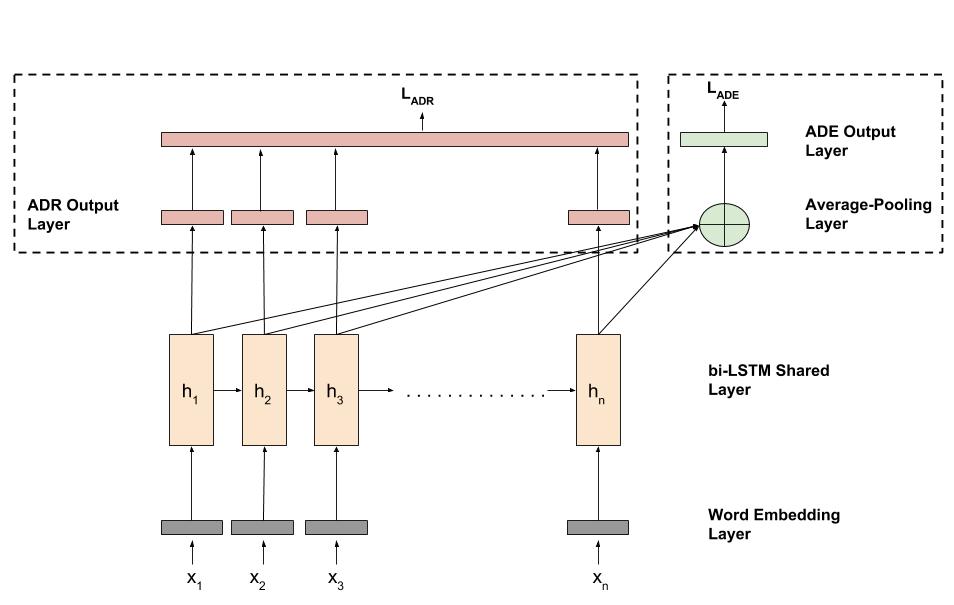}\label{mtl-dia}
  \caption{Network Architecture for the Multi-Task Learning Model to Combine the ADR Extraction and ADE Detection Tasks}
  \label{fig:boat1}
\end{figure}

\noindent \textbf{Multi-Task Learning Model:} The MTL model architecture is illustrated in Fig.~\ref{fig:boat1}. The bi-LSTM transducer acts as the common (shared) layer between both tasks, thus receiving gradient updates from both. The network then bifurcates to task specific layers as seen in the dotted region in the figure. 

The training algorithm is illustrated in Algo.~\ref{alg:tml-adr}. 
To enhance the performance of the main task, we employ the following strategy for training. Since our main task of interest is ADR extraction, the number of parameter updates for this task are fixed to be $N$ (number of training examples for ADR / batch size for ADR) for each epoch. Let $M$ denote the ratio (number of training examples for ADE / batch size for ADE). To compensate for the likely difference in the number of training examples for the ADE task, for each parameter update of the ADR task, $\alpha=\frac{M}{N}$ parameter updates are performed for ADE. 

The MTL setup can also be viewed as an iterative process where each iteration contains two steps. The first step is the detection of an adverse drug event and the second step involves its extraction. We claim that the sharing of the network between the two tasks helps in boosting performance of our main task. We validate this claim in the experiments section.

\subsection{Joint Multi-Task Learning}
\label{sec:jointmtl}
Training a good supervised model for pharmacovigilance need high quality labeled datasets, annotated by domain experts. Getting large datasets labeled by medical domain experts is both time consuming and cost-inefficient. In this section, we discuss a method which can automatically generate auxiliary task dataset in our context, in order to build a MTL pharmacovigilance system. While we discuss the method in the context of pharmacovigilance, it can be applied to other domains too, due to its generic nature. Specifically, we use semi-supervised learning and weakly-supervised learning to augment the main task dataset and generate the auxiliary task dataset respectively. We also present a joint MTL model learned using the data generated using weak-supervision.

\begin{algorithm}[t!]
\scriptsize
\caption{\scriptsize Weakly Supervised Auxiliary Task Dataset Generation}\label{algo2}
\hspace*{\algorithmicindent} \textbf{Input} \hspace{0.3mm}  $U$: Large collection of unlabeled tweets \\
\hspace*{\algorithmicindent} \hspace*{\algorithmicindent} \hspace*{\algorithmicindent}   $\tau$ : threshold for self-training \\
\hspace*{\algorithmicindent} \hspace*{\algorithmicindent} \hspace*{\algorithmicindent}   $D_{ADR}$ : Labeled dataset for ADR task \\
\hspace*{\algorithmicindent} \textbf{Output} \hspace{0.3mm} New labeled datasets $D^{\prime}_{ADR}$ and $D^{\prime}_{ADE}$
\begin{algorithmic}[1]
\State Initialize model parameters, $\theta^{0}$  for bi-LSTM transducer randomly.
\State $T$ $\gets$ $D_{ADR}$\label{algo2:step1}
\While{\textit{(stopping criteria is not met)}}
\State bi-LSTM($\theta^{t}$) = finetune bi-LSTM($\theta^{t-1}$) minimizing $L_{\text{ADR}}$ on $T$\label{algo2:line4} 
\For{$i\gets 1, \vert U \vert$}
\If{$score$($U_i$) $ \ge$ $\tau$ }\label{algo2:line6}
\State $T\gets T \cup U_i$\label{algo2:line7}
\State $U\gets U - U_i$
\EndIf
\EndFor
\EndWhile \label{algo2:step2}\\
$U \gets$ re-sample large pool of unlabeled tweets \label{algo2:step3}\\
$D^{\prime}_{ADR} \gets \phi$ , $D^{\prime}_{ADE} \gets \phi$
\For{$i\gets 1, \vert U \vert$}
\If{$score$($U_i$) $ \ge$ $\tau$ }
\State $D^{\prime}_{ADR} \gets D^{\prime}_{ADR} \cup U_i$\label{algo2:line13}
\State $D^{\prime}_{ADE} \gets D^{\prime}_{ADE} \cup \{U_i, 1\}$\label{algo2:line14}
\Else
\State $D^{\prime}_{ADE} \gets D^{\prime}_{ADE} \cup \{U_i, 0\}$\label{algo2:line15}
\EndIf
\State $U\gets U - U_i$
\EndFor\label{algo2:step4}

\end{algorithmic}
\end{algorithm}

\subsubsection{Weakly Supervised Auxiliary Task Dataset Generation}
Algo.~\ref{algo2} outlines the method to automatically generate auxiliary task dataset (ADE detection in our case). The first stage in the process is to augment the existing training data with a larger dataset generated using semi-supervised learning. Semi-supervised learning can leverage large unlabeled dataset to assist the supervised learning model. We choose self-training~\cite{fralick1967learning}, as the method for semi-supervised learning for this task (Line~\ref{algo2:step1} to~\ref{algo2:step2} of Algo.~\ref{algo2}), mainly because of its simplicity and effectiveness in solving various NLP and IR tasks~\cite{iosifidis2017large,vieira2015self}.

At each step of self-training, the bi-LSTM transducer is trained on the updated training dataset $T$ (Line~\ref{algo2:line4} of Algo.~\ref{algo2}). Note that bi-LSTM's parameters are re-used from the previous iteration. Each sample from the unlabeled example pool is scored using a scoring function computed as follows. First, the current transducer is used to decode/infer output label distribution for each word in the unlabeled sample. For each word in the output sequence, we simply choose the output label which has the maximum probability. We filter out the data sample if the transducer does not output even a single ADR label for any word in the sample. If there is at least one word labeled as ADR, we compute the score for the sample as the multiplication of the ADR probabilities for the ADR-labeled words in the sample normalized by the number of ADR words. If this confidence score of the sample is greater than some pre-defined threshold $\tau$, the sample is added to the training data along with its output labels as generated by the transducer (Line~\ref{algo2:line6} and~\ref{algo2:line7}). 

The next stage is the generation of ADE task dataset (Line~\ref{algo2:step3} to~\ref{algo2:step4}). The pool of unlabeled examples is re-sampled to avoid overlap with the previously used pool. Each data sample from the unlabeled pool is scored using the scoring function defined previously. If the confidence score is greater than $\tau$, the sample is added to the ADR dataset with the decoded labels and it is also added to the ADE dataset with a label of 1 (Lines~\ref{algo2:line13} and~\ref{algo2:line14}). Since this sample's confidence score is greater than a threshold, which indicates with high confidence that it has an ADR mention, it is safe to assume that the sentence has an ADE, thereby assigning it a label of 1. In the other case, due to the low confidence score of the sample, it is assigned a label of 0 for ADE (Line~\ref{algo2:line15}).

\subsubsection{Joint Multi-Task Learning Formulation}
Algo.~\ref{algo2} produces training datasets $D^{\prime}_{ADR}$ and $D^{\prime}_{ADE}$ as the output. We use these to define a joint MTL model as follows. In the dataset, for each example we have two labels, an output label sequence for ADR and a binary label for ADE. We define the joint loss function using  a linear combination of the loss functions of the two tasks as $L = \lambda \cdot \mathbbm{I}{[y_{\text{ADE}} == 1]} \cdot L_{\text{ADR}} + (1 - \lambda) \cdot L_{\text{ADE}}$ where $\lambda$ controls the contribution of losses of the individual tasks in the overall joint loss. $\mathbbm{I}{[y_{\text{ADE}} == 1]}$ is an indicator function which activates the ADR loss only when the corresponding ADE label is 1, since we do not want to back-propagate ADR loss when the corresponding ADE label is 0, which is intuitive by definition.

\section{Experiments}\label{experiments}
In this section we discuss the datasets used, implementation details, experimental results and some qualitative analysis. 

\subsection{Datasets}
The statistics of the datasets are presented in Table~\ref{tab:stats}.
\begin{itemize}
\item We use the Twitter dataset, \textit{Twitter ADR} described in \cite{cocos2017deep}. It contains 957 tweets posted between 2007 and 2010, with mention annotations of ADR and some other medical entities. Due to Twitter's license agreement, authors released only tweet ids with their corresponding mention span annotations. At the time of collection of the original tweets using Twitter API, we were able to collect only 639 tweets. 
\item We use the second Twitter dataset, \textit{TwiMed} described in \cite{alvaro2017twimed}. It contains 1000 tweets with mention annotations of Symptoms from drug (ADR) and other mention annotations posted in 2015. Due to Twitter's license agreement, we were able to extract 663 tweets only. 
\item For the ADE detection task, we use the Twitter dataset \textit{Twitter ADE} released as part of a Health application shared task\footnote{https://healthlanguageprocessing.org/}. The dataset consists of 13829 tweets annotated with a label of 1 or 0 indicating the presence or absence of an adverse drug event respectively.
\item For the unlabeled tweets used for semi-supervised learning, we collected tweets using the keywords as drug-names and ADR lexicon publicly available\footnote{http://diego.asu.edu/downloads}. This filtering step ensures that all collected tweets have at least one drug-name occurrence and one ADR phrase. The tweets were posted in 2015. 
\end{itemize}

\begin{table*}
\scriptsize
\centering
 \begin{tabular}{|c | c | c | c | c|} 
 \hline
 Dataset & No. tweets & No. ADR Words& Pos. ADE & Neg. ADE   \\  
 \hline
 Twitter ADR & 639 & 1,526 & - & - \\ 
 \hline
 TwiMed & 663 & 1,091 & - & - \\
 \hline
 Twitter ADE & 13,829 & - & 1,206 & 12,623 \\
 \hline
 Unlabeled Tweets & 4,61,522 & - & - & - \\
 \hline
\end{tabular}
\caption{Dataset Statistics}\label{tab:stats}
\end{table*}

\subsection{Implementation Details}
For implementation of the model, we use the popular python deep learning toolkits Keras\footnote{https://keras.io/} and Tensorflow\footnote{https://www.tensorflow.org/}. 

\noindent \textbf{Text Pre-processing:} As part of text pre-processing, we normalized all HTML links and USER mentions to the tokens ``$\langle$LINKS$\rangle$'' and ``$\langle$USER$\rangle$'' respectively. We limit the vocabulary size to 40k most frequent words in case of semi-supervised learning based MTL task. We also remove all mentions of special characters and emoticons from the tweet. For each method, the tweet length is padded to the maximum length from the corpus.

\noindent \textbf{Hyper-parameter settings:} We kept the hyper-parameter setting for the bi-LSTM transducer similar to the one reported in \cite{cocos2017deep}. Word2Vec embeddings trained on a large generic tweet collection with a dimension of 400 \cite{godin2015multimedia} are used as input to the transducer. The hidden unit dimension ($d_h$) is set to 500. The number of output units ($d_l$) is 4. We use adam \cite{kingma2014adam} as optimizer with number of epochs set to 10 for all methods. The batch-size for the ADR and ADE tasks are set to 8 and 32 respectively for the MTL method. For the semi-supervised learning method in the weak-supervision part, the batch-size for the ADR task is set to 64 with the confidence threshold value empirically set to 0.5. The stopping criteria for the self-training kicks in when the number of iterations reaches 5 or if the unlabeled tweets pool is exhausted, whichever occurs first. For the joint MTL method, the $\lambda$ is empirically set to 0.8. The learning rate for all methods is set to 0.001.   

\subsection{Results}
\begin{table}[t!]
\centering
\scriptsize
 \begin{tabular}{|l | c | c | c |} 
 \hline
 \textbf{Method} & \textbf{Precision} & \textbf{Recall} & \textbf{F1-score}   \\
 \hline
 Baseline \cite{cocos2017deep} &  0.7067 $\pm$ 0.057 & 0.7207 $\pm$ 0.074 & 0.7102 $\pm$ 0.049 \\
 Baseline with adam &  0.7065 $\pm$ 0.058 & 0.7576 $\pm$ 0.083 &	0.7272 $\pm$ 0.051 \\
 KB-Embedding Baseline \cite{DBLP:conf/eacl/StanovskyGM17} & 0.7171 $\pm$  0.058 &	0.7713 $\pm$ 0.091 &	0.7397 $\pm$ 0.055 \\
 \hline 
 Self-training & 0.6999 $\pm$ 0.047 &	0.8304 $\pm$ 0.039 &	0.7588 $\pm$ 0.039 \\
 Joint MTL (Section~\ref{sec:jointmtl})& 0.7177 $\pm$ 0.027	& \textbf{0.8482 $\pm$ 0.068} &	0.7770 $\pm$ 0.043 \\
 MTL (Section~\ref{sec:mtl})& \textbf{0.7569 $\pm$ 0.044} &	0.8386 $\pm$ 0.078 &	\textbf{0.7935 $\pm$ 0.045} \\
  \hline 
\end{tabular}
\caption{Experimental Results for Twitter ADR dataset (along with Std. Deviation)}\label{tab:results1}
\end{table}

The results of various methods are presented in Tables~\ref{tab:results1} and~\ref{tab:results2} for the Twitter ADR  and TwiMed datasets respectively. For the ADR task, to encode the output labels we use the IO encoding scheme where each word is labeled with one of the following labels: (1) I-ADR (ADR mention), (2) I-Other (mention category other than ADR), (3) O, (4) PAD (padding token). Since our entity of interest is ADR, we report the results on ADR only. An example tweet annotated with IO-encoding is as follows. ``\textit{@BLENDOS\textsubscript{O} Lamictal\textsubscript{O} and\textsubscript{O} trileptal\textsubscript{O} and\textsubscript{O} seroquel\textsubscript{O} of\textsubscript{O} course\textsubscript{O} the\textsubscript{O} seroquel\textsubscript{O} I\textsubscript{O} take\textsubscript{O} in\textsubscript{O} severe\textsubscript{O} situations\textsubscript{O} because\textsubscript{O} weight\textsubscript{I-ADR} gain\textsubscript{I-ADR} is\textsubscript{O} not\textsubscript{O}  cool\textsubscript{O}}''. For performance evaluation we use approximate-matching \cite{tsai2006various}, which is used popularly in biomedical entity extraction tasks \cite{cocos2017deep,nikfarjam2015pharmacovigilance}. We report the F1-score, Precision and Recall computed using approximate matching as follows.

\small
\begin{equation}
\text{Precision} = \frac{\text{\#ADR approximately matched}}{\text{\#ADR spans predicted}}, \text{Recall}= \frac{\text{\#ADR approximately matched}}{\text{\#ADR spans in total}}
\end{equation}
\normalsize
The F1-score is the harmonic-mean of the Precision and Recall values. All results are reported using 10-fold cross-validation along with the standard deviation across the folds.

Our baseline methods are bi-LSTM transducer \cite{cocos2017deep} with traditional word embeddings and the current state-of-the-art bi-LSTM transducer which used traditional word embeddings augmented with knowledge-graph based embeddings \cite{DBLP:conf/eacl/StanovskyGM17}. 

For both the datasets, it should be noted that Cocos et al. \cite{cocos2017deep} used RMSProp as an optimizer, and since we are using adam for all our methods, so for a fair comparison we also report the baseline results with adam. The corresponding results are reported in the first two rows of both the tables. It is clear that re-implementation with adam optimizer enhances the performance, which is consistent with the general consensus around adam optimizer. 

The KB-embedding baseline \cite{DBLP:conf/eacl/StanovskyGM17} replaces word embeddings of the medical entities in the sentence with the corresponding embeddings learned from a knowledge-base. The corresponding results can be seen in row 3 of the tables. It is clear that adding KB-based embeddings enhances the performance over the baseline, due to the external knowledge added in the form of KB embeddings. 

The results for our methods are presented from row 4 onwards. We first discuss the results from our joint MTL method. Since the joint MTL method involves self-training as its first step followed by the joint modeling, we also report results using self-training alone. Results from self-training are reported in row 4 in the tables. The self-training based method outperforms the KB-based method, which shows that addition of a large unlabeled corpus in the model improves the performance. 

Addition of another task on top of unlabeled data and modeling it in a joint MTL setting further improves the performance. Finally, the results from the MTL method using actual ADE task dataset are presented in the last row of both the tables. It can be seen that the MTL method significantly outperforms baseline methods in terms of F1-score. These results validate our initial hypothesis that sharing two similar tasks of ADR extraction and ADE detection helps the model generalize better.

\begin{table}[t!]
\centering
\scriptsize
 \begin{tabular}{|l | c | c | c |} 
 \hline
 \textbf{Method} & \textbf{Precision} & \textbf{Recall} & \textbf{F1-score}   \\  
 \hline
 Baseline \cite{cocos2017deep} &  0.6120 $\pm$  0.116 &		0.5149 $\pm$  0.099 &		0.5601 $\pm$  0.100 \\
 Baseline with adam &  0.6281 $\pm$ 0.094	&	0.5614 $\pm$ 0.110	&	0.5859 $\pm$ 0.079 \\
 KB-Embedding Baseline \cite{DBLP:conf/eacl/StanovskyGM17} & 0.5960 $\pm$ 0.081 &		0.6144 $\pm$ 0.068	&	0.6042 $\pm$ 0.060 \\
 \hline 
 Self-training & 0.5717 $\pm$ 0.056 &	0.7141 $\pm$ 0.082 &	0.6332 $\pm$ 0.057 \\
 Joint MTL (Section~\ref{sec:jointmtl}) & 0.5675 $\pm$ 0.049 &	\textbf{0.7384 $\pm$ 0.079} &	0.6401 $\pm$ 0.051 \\
 MTL (Section~\ref{sec:mtl})& \textbf{0.6656 $\pm$ 0.083}	&	0.6380 $\pm$ 0.077	&	\textbf{0.6482 $\pm$ 0.065} \\
  \hline  
\end{tabular}
\caption{Experimental Results for TwiMed dataset (along with Std. Deviation)}\label{tab:results2}
\end{table}

\section{Qualitative Analysis}\label{analysis}

In this section, we aim to answer the following research questions.
\begin{itemize}
\item \textbf{Q1:} What is the effect of the auxiliary task dataset's size on the performance of the MTL method?
\item \textbf{Q2:} What is the effect of size of the unlabeled corpus on the performance of self-training and joint-MTL method?
\item \textbf{Q3:} What is the effect of adding more depth to the bi-LSTM transducer on the MTL method's performance?
\end{itemize}

\begin{figure}[!htb]
\minipage{0.32\textwidth}
\includegraphics[width=\columnwidth]{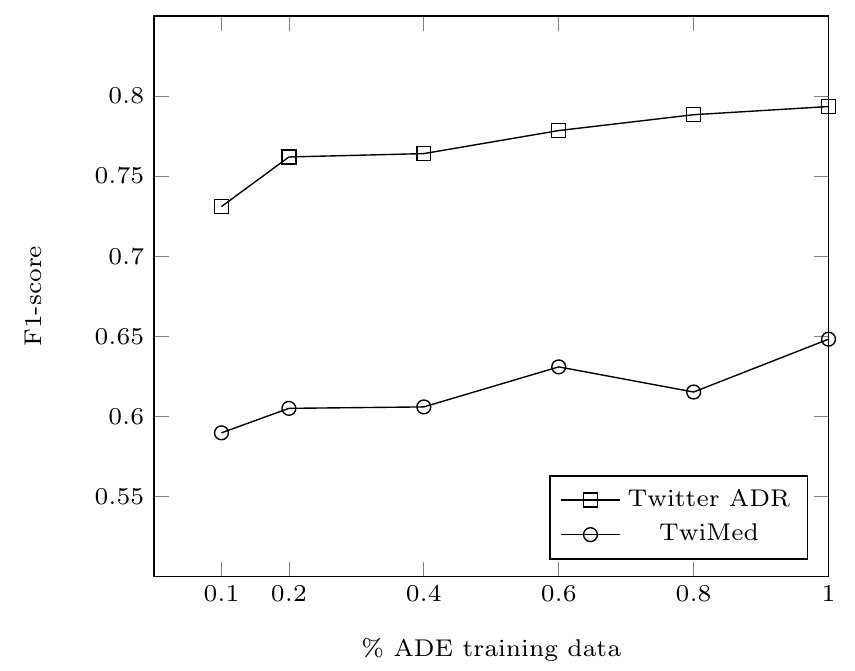}%
\caption{Performance variation with \% ADE data-size}%
\label{fig1}%
\endminipage\hfill
\minipage{0.32\textwidth}
\includegraphics[width=\columnwidth]{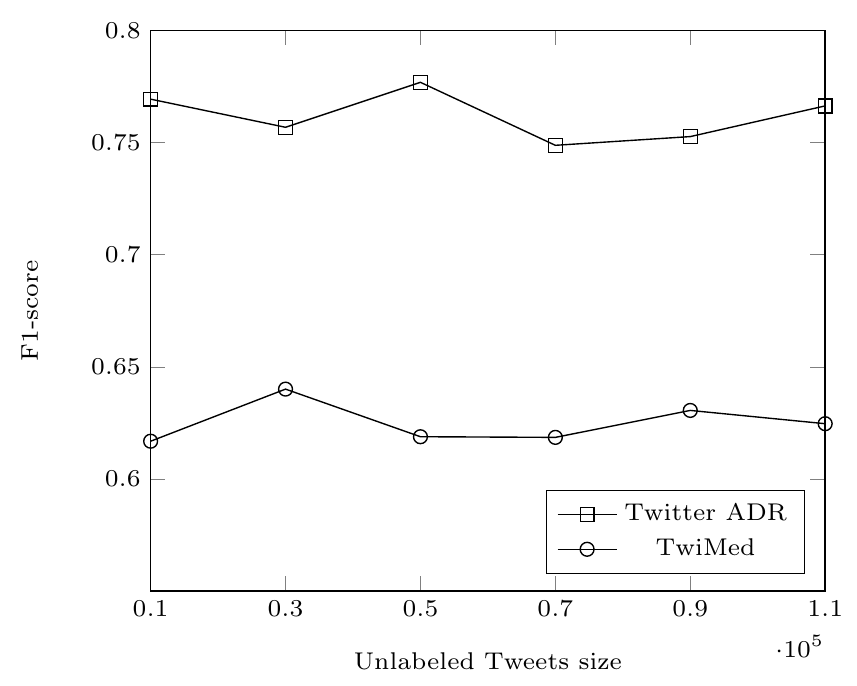}%
\caption{Performance variation with unlabeled data-size}%
\label{fig2}%
\endminipage\hfill
\minipage{0.32\textwidth}
\includegraphics[width=\columnwidth]{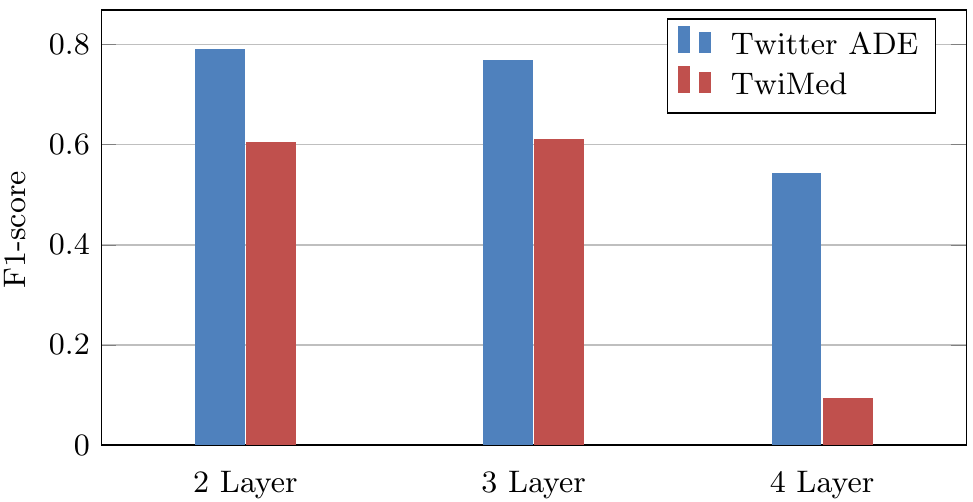}%
\caption{Performance variation with stacking bi-LSTM layers}%
\label{fig3}%
\endminipage
\end{figure}

To answer the first question, we perform MTL experiments with varying ADE dataset size. The results are presented in Figure \ref{fig1}. F1-score has a clear correlation with the percentage training size for the ADE task. As the ADE dataset size is increased, the F1-score also increases monotonously. Similar trend is observed for both the datasets. It clearly indicates the importance of the auxiliary task in the MTL setting.  

To answer the second question, we perform joint MTL experiments with varying unlabeled data size. Results are presented in Figure \ref{fig2}. The results are fairly flat for both the datasets as the unlabeled data size increases. This clearly indicates that our joint MTL method is robust to the size of unlabeled data. It also indicates that our method works well even with a small seed set of unlabeled data-points too.

Results with varying representation capacity of bi-LSTM transducer are presented in Figure \ref{fig3}. It is clear that the performance degrades as more bi-LSTM layers are stacked on top of the original model. We suspect that this might be the case due to limited manually annotated training data present.

\section{Conclusions}\label{conclusion}
In this paper, we proposed two multi-task learning based methods to tackle the problem of labeled data scarcity for adverse drug reaction mention extraction task. Our first method uses adverse drug event detection as an auxiliary task, and demonstrates superior results in comparison to performing the ADR extraction task independently. The second proposed method is a novel joint MTL method, which uses semi-supervised and weakly-supervised learning to automatically generate ADE detection task dataset and then uses the datasets in a novel joint-MTL setting where both tasks are simultaneously modeled. We analyzed the method on two popular ADR extraction datasets, and it demonstrates superior results as compared to the state-of-the-art methods in ADR extraction. 

\end{document}